 \documentclass[manuscript]{aastex}

\newcommand{\gtwid}{\mathrel{\raise.3ex\hbox{$>$\kern-.75em\lower1ex\hbox{$\sim$}}}}
\newcommand{\ltwid}{\mathrel{\raise.3ex\hbox{$<$\kern-.75em\lower1ex\hbox{$\sim$}}}}

\shorttitle{A Hidden Nucleus in Cygnus A, but Not in M87}
\shortauthors{Whysong and Antonucci}

\begin{document}


\title{A Hidden Nucleus in Cygnus A, but Not in M87}


\author{D. Whysong and R. Antonucci}
\affil{Physics Department, University of California, Santa Barbara, CA 93106}







\begin{abstract}
Historically the narrow line radio galaxies were thought to be
intrinsically nonthermal, and without significant accretion.
When the Unified Model came along the compelling observational
motivation for this lost some force: some are found to be
{\em hidden}\/ broad line objects, and in principle that could be the
case with all of them.
    The clear sign of a hidden quasar is a normal quasar spectrum
in polarized (scattered) light.  However that test requires a suitably
placed ``mirror."  A more robust test is the high predicted mid-IR core
luminosity reradiating from the obscuring matter.  Cygnus A has this
component, but M87 does not.
\end{abstract}

Subject headings: galaxies: active --- galaxies: individual (M87, 3C 405) ---
galaxies: nuclei --- infrared: galaxies




\section{Introduction}

  Now that the existence of supermassive black holes in AGN seems fairly
secure, perhaps the next most fundamental questions are the source of
energy and nature of the accretion flow in the various classes of objects.
Historically (see Begelman, Blandford and Rees 1984 for an early review)
it was thought that the optical/UV continuum (the ``Big Blue Bump") in quasars
(hereafter: and broad line radio galaxies)
was thermal radiation from some sort of cool optically thick accretion
flow.  Radio galaxies didn't show this component, so were posited to be 
``nonthermal AGN" with hot radiatively inefficient accretion at a very
low rate, with the jet power deriving from the hole rotation rather than
accretion, perhaps via the Blandford-Znajek (1977) mechanism.

  One Fanaroff-Riley II (edge-brightened, very luminous) radio galaxy,
3C234, was shown in 1982 to have quasar features
(broad permitted emission lines) in polarized light (Antonucci 1982,
1984;  Antonucci 1993 for a review).  Thus it does have
the ``thermal" Big Blue Bump, which is only visible via scattering.
Many other examples have been shown subsequently (e.g.\ Hines and Wills 1993,
Young et al.\ 1996).
Some invocations of the Unified Model postulated that this was also true
of the FR II (powerful, edge-brightened) class generally (e.g. Barthel 1989).
However, it is still contentious how those FR IIs with weak and/or low
ionization narrow emission lines fit in (Antonucci 2001, Singal 1993, Laing 1994, 
Gopal-Krishna et al.\ 1996).
The situation for the FR I galaxies, almost all of which have undetectable
or low-ionization emission lines, is even less clear.
Where there is no observable high-ionization narrow emission line region
present, there is no {\em a priori}\/ evidence for the presence of
a quasar.  These could still have a quasar nucleus, but any narrow line
region would need to be mostly obscured as well.  That seems to be the case
for NGC4258 (Wilkes et al.\ 1995, Barth et al.\ 1999).

{\em Many}\/ FR II radio galaxies fall into the weak, low-ionization category. 
{\em Most}\/ of the FR I radio galaxies do as well.  Among these objects
a majority show optical/UV point sources, while a significant minority 
do not (Chiaberge et al.\ 1999, 2000).
The galaxies known from spectropolarimetry to harbor hidden thermal nuclei
generally do not show point sources, but just extended emission from
the mirror.  (3C234 is an exception, though it also has some extended flux.)
  How can we tell whether or not a hidden nucleus is present?
Many hidden AGN that are not Compton-thick have been discovered with X-ray
observations (Antonucci 2001); NGC4945 is a spectacular example (Madejski et al 2000).
Detection of broad lines in the polarized flux spectra indicates
robustly the presence of a hidden normal quasar (or broad line radio
galaxy).  But this test requires fortuitous placement of a natural
scattering mirror.  A much more complete test is to measure the
reradiation from nuclear hot and warm dust --- which almost necessarily
accompanies any hidden AGN.  Unlike spectropolarimetry this method
measures the ``waste heat" and so provides an estimate of the hidden
luminosity.  We're half-way through a major survey of the 3C radio galaxies in
the mid-IR with Keck, and thought the results for M87 and Cygnus A were worth
showing now.

\section{Observations}

\subsection{3C 405 (Cygnus A)}

We present a diffraction limited mid-IR image of the nuclear source in
3C 405 (Cyg A), obtained with the Long Wavelength Spectrometer (LWS)
instrument at the Keck telescope\footnotemark[1].  All data
were taken with the 11.7$\mu$m filter, which has an $\sim1\ \mu$m
bandpass from 11.2 to 12.2 $\mu$m.

The nucleus of 3C 405 was imaged at 11.7$\mu$m with Keck/LWS on 1999 September
30. The chop/nod throw was set to 10{\tt "} in order to allow imaging
of larger scale extended structure; this places the negative images off
the chip, which has a 10{\tt "} field. We do not report on structures larger
than the 10{\tt "} chop distance. The images were dithered in a 5 position
box pattern, 2{\tt "} to a side, with 53.1 seconds on-source for the positive
image per dither position. The entire 5-position exposure was repeated three
times, for a total on-source time of 796.6 seconds.

Data were processed by subtracting all background chop/nod frames, shifting
each dithered image to the correct position, and coadding all dither images.
Morphology is extended, with structure to the east and southeast of the nucleus
(Fig.~1).  The standard star was alpha Ari\footnotemark[2], with a FWHM of 0.27{\tt "}.

$$\vbox{\halign{#\hfill&\qquad\hfill#\cr
\noalign{Table 1: LWS photometry results for Cygnus A:}
 aperture diameter  &     flux (mJy)\cr
     (arcsec)\cr
      0.64     &              44\cr
      0.96     &              71\cr
      1.28     &              93\cr
      1.60     &             111\cr
      1.92     &             122\cr
      2.56     &             139\cr
      3.20     &             152\cr
}}$$

For comparison, the IRAS (large aperture) data for Cygnus A are listed in 
Table 2.

$$\vbox{\halign{#\hfill&\qquad\hfill#\cr
\noalign{Table 2: (Impey and Neugebauer 1988):}
    12 $\mu$m &  S = 144 +/- 5 mJy\cr
    25 $\mu$m &  S = 870 +/- 5 mJy\cr
    60 $\mu$m &  S = 2908 +/- 13 mJy\cr
}}$$

There is Galactic emission contamination at longer wavelengths.

Our 11.7$\mu$ image and a partial spectral energy distribution are shown
in Figs.~1 and 2.

\subsection{3C 274 (M87)}

Similar imaging data were obtained for M87 on 2000 January 18.  The instrument
was observed in chop-nod mode using a small 3.5{\tt "} amplitude so as to keep
both the positive and negative images on the CCD chip.

The same dither pattern was used, with 96 seconds per dither each for positive
and negative (background).  Unfortunately, due to a loss of guiding, the
positive nucleus image was only fully imaged in one dither frame.

Beta Andromedae and Mu Ursa Majoris were used as standard stars for photometric
calibration, yielding a flux scale of 0.0874 and 0.0931 mJy/(ADU/s) and FWHM
of 0.33{\tt "} and 0.31{\tt "} respectively. Applying this photometric
calibration to the unresolved nuclear component in M87 results in a flux of 13
+/- 2 mJy. The uncertainty is dominated by systematic errors in the background
subtraction; we conservatively adopt a value of 15 mJy. A synthetic aperture of
0.96 arcsec was used, but the source is unresolved so the flux is insensitive
to the aperture.

The IRAS fluxes are listed here:
(Moshir et al.\ 1990).

$$\vbox{\halign{#\hfill&\qquad#\hfill\cr
\noalign{Table 3.}
    12 $\mu$m &  S =  231 +/- 37 mJy\cr
    25 $\mu$m &  S $<$  241 mJy\cr
    60 $\mu$m &  S =  393 +/- 51 mJy\cr
}}$$

\section{Discussion}

\subsection{Cygnus A}

This is a very powerful FR II radio galaxy at a redshift of 0.056.  It has
strong high-ionization narrow lines, suggestive of a hidden AGN.  A broad Mg II
2800 emission line is detectable in total flux (Antonucci et al.\ 1994).  That
line may or may not be highly polarized, and thus scattered from a hidden
nucleus.  Several detailed papers report spectroscopic and spectropolarimatric
data (Goodrich and Miller 1989, Tadhunter et al.\ 1994, Shaw and Tadhunter
1994, Vestergaard and Barthel 1993, Stockton et al.\ 1994;  see also Tadhunter
et al.\ 2000 and Thornton et al.\ 1999 - and there are several others),
culminating in Ogle et al.\ 1997, which shows an extremely broad H-alpha line
in polarized flux.  It is virtually invisible in total flux.

A nuclear point source in the near-IR was noted by Djorgovski et al.\ (1991),
but they don't seem to have considered hot dust emission for this excess over
the extrapolation of the optical light.

A powerful hidden nucleus should manifest a mid-IR dust luminosity much larger
than the observed optical luminosity.  However for this object and M87 (and
virtually all others!)  the only IR data available were taken with very large
beam sizes\footnotemark[3].  We isolate the core much better with the $\sim0.3\ \rm arcsec =
\sim1.1$ kpc resolution provided by the Keck telescopes, and find a nuclear
flux of $\sim 71$mJy.  An uncertainty here derives from the extended emission,
but flux as a function of aperture size does flatten out for apertures larger
than the seeing disk, so the 0.96{\tt "} measurement should be approximately
correct (see Table 1).  However, we can't be sure from this observation alone
that the emission is on pc scales.  Since the emission is powerful and at the
relatively short wavelength of 11.7$\mu$, it is very likely that this comes
from nuclear dust rather than a starburst.

The IRAS (large aperture) dust spectrum is quite cool, suggesting a large
starburst contribution.  Extended emission is seen in our image at 11.7$\mu$
but for the present purpose we want the nuclear flux.  The core can't be
exactly separated from the extensions (see Table 1), but we can estimate around
60 mJy for an unresolved core.  Fig.~1 shows the Cyg A 11.7$\mu$ image, and
Fig.~2 shows a partial spectral energy distribution.  The nuclear luminosity
vLv at 11.7$\mu$ is 10 times higher than that at 0.5$\mu$.  The latter
wavelength needs two roughly canceling corrections:  subtraction of optical
light from the host galaxy, and dereddening (Ogle et al.\ 1997).  

The conclusion is simple and expected from prior evidence:  Cyg A has
a moderately powerful hidden nucleus.  The redshift is 0.056, and we
adopt a Hubble constant of 75 km/s Mpc.  This implies a vLv luminosity 
at 11.7$\mu$ of $9.2\times 10^{43}$ erg/sec
If the intrinsic SED is similar
to those of PG quasars (Sanders et al.\ 1989), the 11.7$\mu$ value
implies a bolometric luminosity of 16.5 vLv(11.7$\mu$)
$\sim 1.5\times 10^{45}$ erg/sec.  
For comparison the jet power is estimated several different ways
(Carilli and Barthel 1996,  Sikora 2001,  Punsly 2001).  The values
are rough, but generally lie in the $\gtwid10^{45}$ ergs/sec range.
This is consistent with the finding that jet power and optical/UV
luminosity are often comparable in double radio quasars (e.g., Falcke 1995).

Thus this is a moderate luminosity broad line radio galaxy with a very
high luminosity radio source.  It has
in fact been inferred already that Cygnus A is an over-achiever 
in the radio (Carilli and Barthel 1996, Barthel and Arnaud 1996).  
This is well explained qualitatively by the fact that it is the only known
FR II radio source in a rich X-ray-emitting cluster.

\subsection{M87}

This radio galaxy is on the FR I-II borderline, both in morphology
and in radio power (Owen et al.\ 2000).  It is one of the majority
of FR I radio galaxies with an optical/UV point source (Chiaberge et al 1999, 
2000). The optical point source is tentatively ascribed to synchrotron
radiation associated with the radio core (see Ford \& Tsvetanov 1999).
However it's not yet known whether the source is highly polarized, and whether
broad emission lines are strong in either total or polarized optical/UV flux.  

Our small $\sim0.3$ arcsec beam isolates the innermost $\sim25$pc in this case,
and the enclosed 11.7$\mu$ flux is 15 mJy, much lower than the large aperture
measurements.  The large-aperture data from the literature leave plenty
of room for waste heat from 
a hidden AGN, but our data do not (Fig.~3).  The mid-IR luminosity
is only comparable to that in the optical.  In fact, much or all of the
mid-IR flux could be synchrotron radiation associated with the 
innermost part of the jet seen in the radio and perhaps optical, so our
measurement should be considered an upper limit to the dust luminosity.

Subject to the caveat that the nuclear dust emission could be entirely too cool
to emit in the near-IR, this rules out a powerful hidden nucleus, and the
actual upper limit on its power is of interest.  The observed 11.7$\mu$ flux
corresponds to vLv = $1.0\times 10^{41}$ erg/sec, for a distance of 15 Mpc.
For the SED for the PG quasar composite of Sanders et al.\ 1989, a bolometric
luminosity of $\sim1.6\times 10^{42}$ ergs/sec (e.g., the templates in Sanders
et al.\ 1989 and Barvainis 1990).  For clues to the nature of the central
engine, we can compare this upper limit on any quasar-like nucleus to the power
in the radio jet.  A lower limit to the jet kinetic luminosity in M87 is
$\sim5\times10^{44}$ erg/sec (Owen et al.\ 2000), so the jet is by far the
dominant channel for energy release.

  The ADAF predictions (Reynolds et al.\ 1996) are for very low IR-optical-UV
luminosities compared with those in the radio and X-ray.  Our 11.7$\mu$ point
lies at a level about equal to the optical value, but certainly our 11.7$\mu$
point may be partially or completely jet emission\footnotemark[4].

\section{Relation to other Radio Galaxies and Conclusion}

Since the M87 optical/UV flux is quite variable (e.g., Tsvetanov
et al.\ 1998; see also the review, Ford and Tsvetanov 1999),
jet synchrotron emission is a possibility.  By correlating the
radio synchrotron core fluxes and the optical point source fluxes
in FR I radio galaxies generally, Chiaberge et al.\ (1999) infer that
the latter are in fact beamed synchrotron sources.
A crucial test of the nature of the optical point sources is
spectroscopy (and polarimetry). We hope to do this with adaptive
optics, excluding most of the starlight that dominates in arcsec
apertures.

Radio lobe emission is fairly isotropic, so it's easy to make lists of radio
galaxies that are nearly unbiased with respect to orientation. The visibility
of optical point sources in {\em most}\/ of the optically dull (weak or low
ionization emission line) galaxies show that there are no tori generally
present which are able to obscure the optical point source (Chiaberge et al.\
1999, 2000). The optical source, whatever it's nature, is tiny --- if they all
vary as M87 does. In that case Chiaberge et al.\ would probably be correct in
concluding that these objects have no hidden nuclei.

However, a large minority of low ionization FR I and FR II radio galaxies show
no point source, and are similar in this way to the AGN with hidden nuclei. In
fact the closest FR I, Cen A, {\em does}\/ have a big molecular torus (see
Fig.~2 of Rydbeck et al.\ 1993!), and substantial evidence for a hidden nucleus
as well (Capetti et al.\ 2000 and references therein; see also Marconi et al.\
2000, Ekers and Simkin 1983, Sambruna et al.\ 2000 for more hidden nuclei in
sources with weak and/or low ionization emission lines). As a working
hypothesis we suppose that the same is true for all those without detectable
optical pointlike nuclei. (Of course sensitivity of the optical/UV
observations also enters in.)

Thus the FR I family is a heterogeneous one, with some contain hidden nuclei
and some not. In fact at least a few are known to be quasar-like from direct
spectroscopy (3C120 is well known; see also Lara et al.\ 1999 and Sarazin et
al.\ 1999).

It is absolutely not the case therefore that the ``nonthermal" model
applies to all optical dull or low ionization radio galaxies, 
or to all FR I galaxies. The FR class has no apparent direct relation
to the mode of energy production, consistent with much recent evidence
that the FR Is behave very much like FR IIs at VLBI scales.

\footnotetext[1]{Instrument reference is available at:
http://www2.keck.hawaii.edu:3636/realpublic/inst/lws/lws.html.}

\footnotetext[2]{A table of photometric standards is available at:
http://www2.keck.hawaii.edu:3636/realpublic/inst/lws/IRTF$\_$Standards.html.}

\footnotetext[3]{Photometry data can be found at: http://nedwww.ipac.caltech.edu.}

\footnotetext[4]{It is unclear to us why the radio points, which fit the model,
are taken as measurements while the (non-fitting) optical fluxes were not.
Also, the Reynolds et al.\ figure apparently uses 3C273 as a ``thermal"
quasar-like template, but that object definitely has a large jet contribution
in the radio and infrared (Robson et al.\ 1993).}


\clearpage


\begin{figure}
\plotone{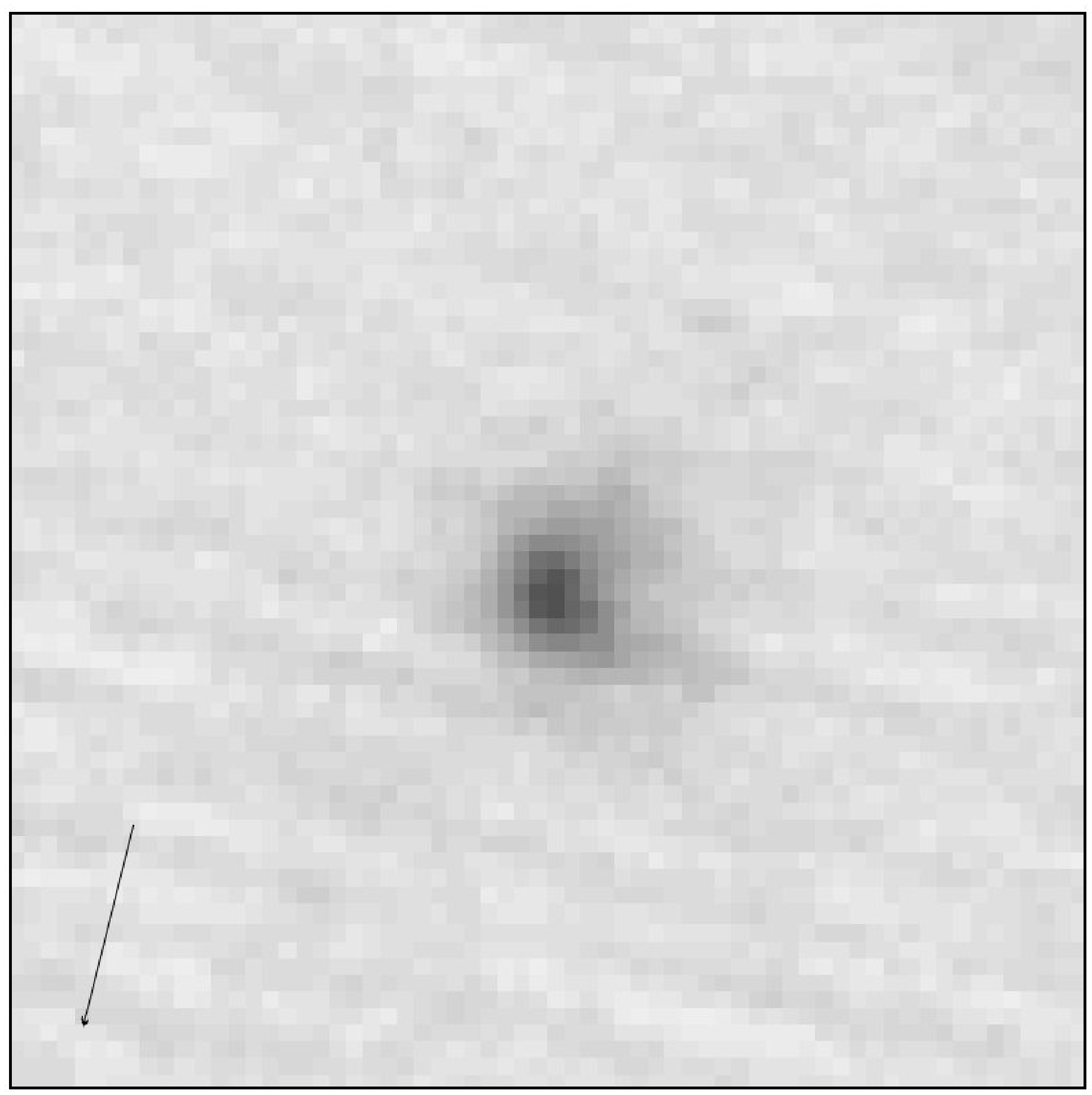}
\caption{Keck LWS 11.7$\mu$m image of Cygnus A. The arrow indicates north and is 1 arcsec long.}
\end{figure}

\clearpage 

\begin{figure}
\plotone{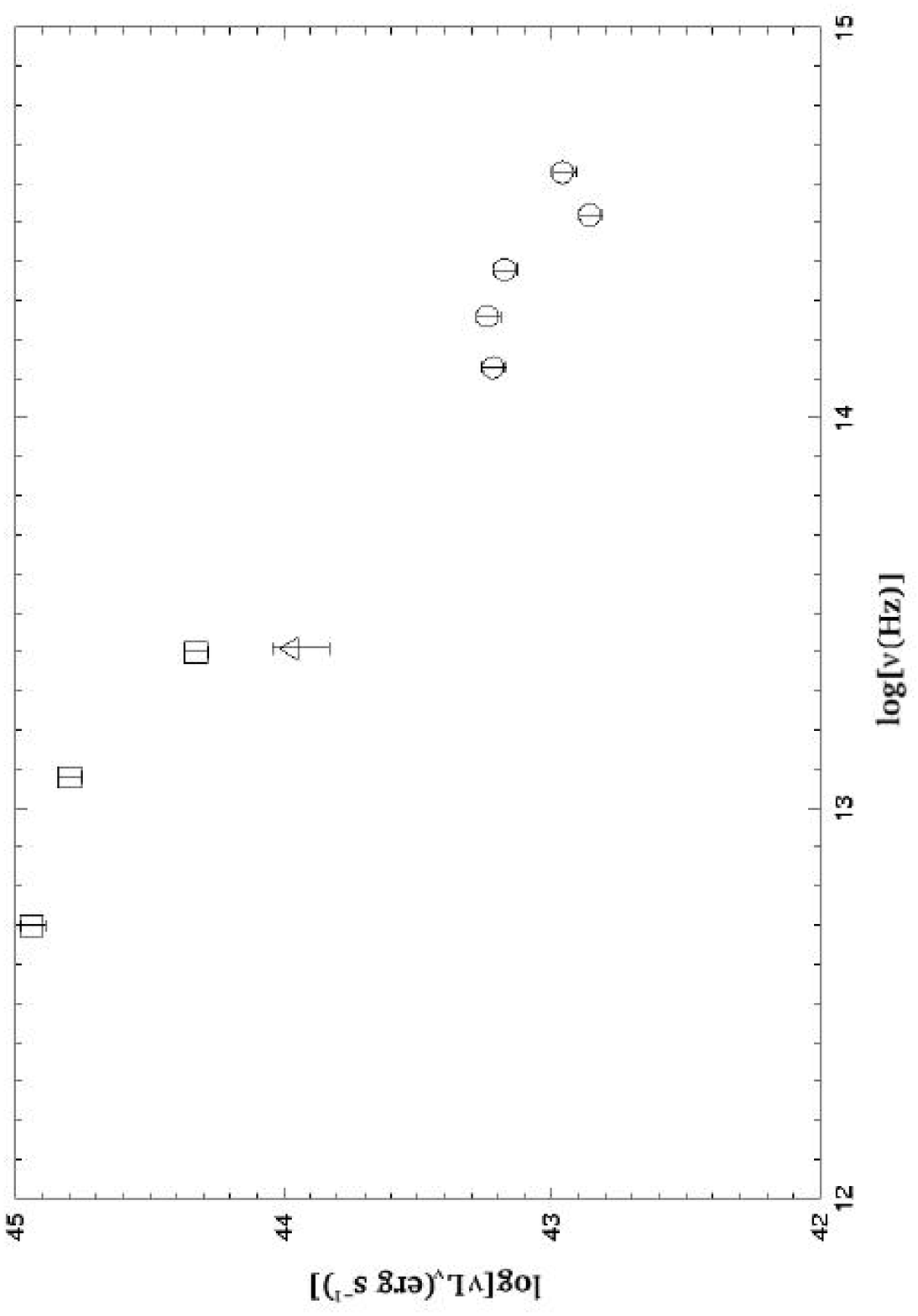}
\caption{Partial core SED for Cygnus A. Data are from IRAS (squares, Impey and
Neugebauer 1988), Keck/LWS (triangle), and Palomar 200 inch (circles, Djorgovski
et al.\ 1991). Errors for the Keck/LWS point represent uncertainty in nuclear
emission due to the extended structure.  Other data are plotted with 10\% error
bars. While measurement errors were much smaller than this, we include these
larger uncertainties due to the different apertures.}
\end{figure}

\clearpage 

\begin{figure}
\plotone{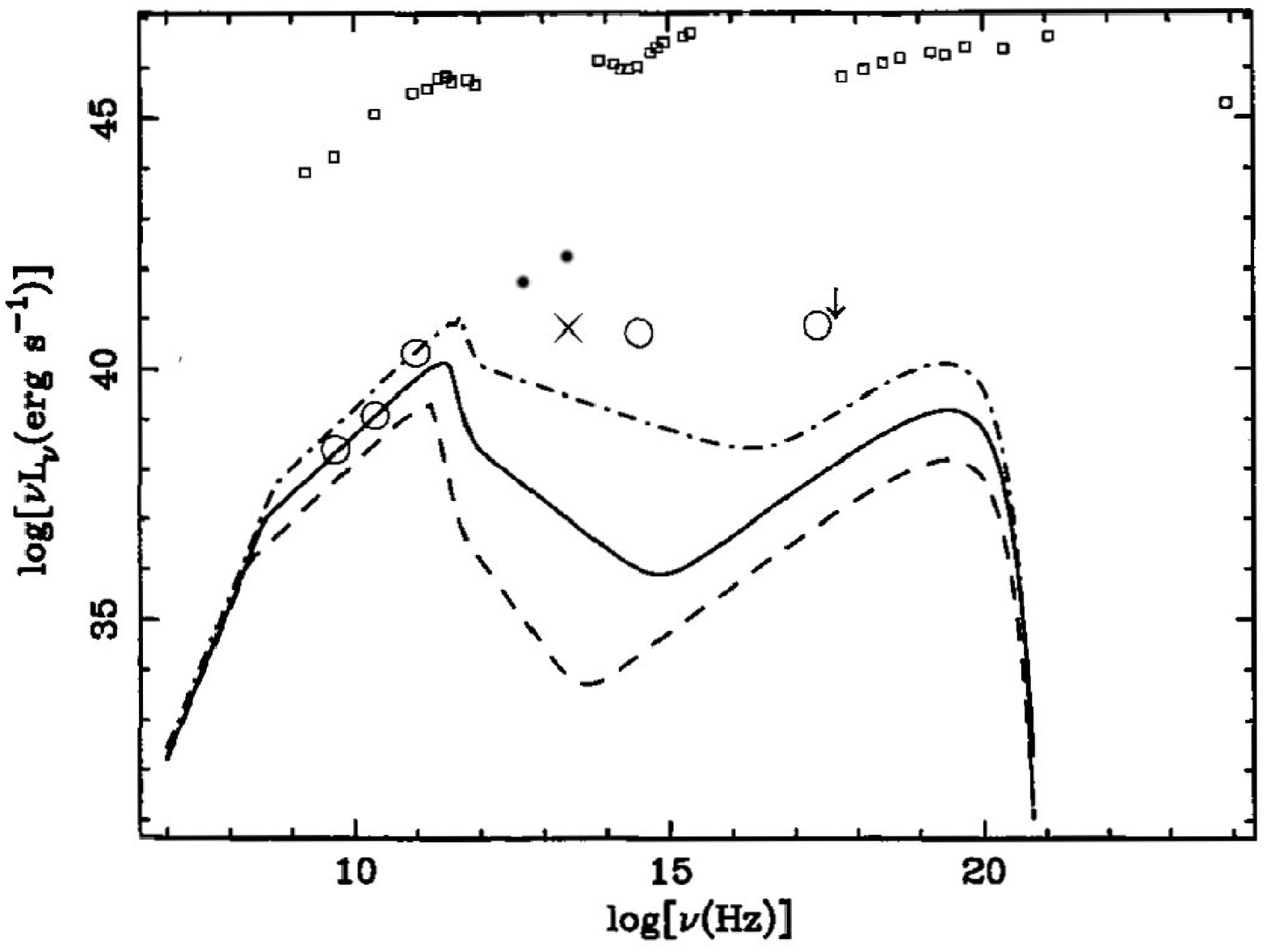}
\caption{Spectral energy distribution for M87 (open circles) and 3C273
(squares), and ADAF models (lines) for various accretion rates (Reynolds et
al.\ 1996). Additional M87 points are from the IRAS faint source catalog
(Moshir et al. 1990) (solid circles) and Keck/LWS (crosses).}
\end{figure}

\clearpage


\begin{thebibliography}{}
\bibitem{ref:1} Antonucci, R 1982 Nature 299, 605
\bibitem{ref:2} Antonucci, R 1993, Ann Rev Astron Astrophys 31, 473
\bibitem{ref:3} Antonucci, R 1984 ApJ 278, 499
\bibitem{ref:4} Antonucci, R 2001, preprint (astro-ph/0103048)
\bibitem{ref:5} Antonucci, R, Hurt, T, and Kinney, A 1994 Nature 371, 313
\bibitem{ref:6} Barth, A J, Hien, H D, Brotherton, M S, Filippenko, A V,
 Ho, L C, van Breugel, W, Antonucci, R, and Goodrich, R 1999 A J 118, 1609
\bibitem{ref:7} Barthel, P 1989 ApJ 336, 606
\bibitem{ref:8} Barthel, P D, and Arnaud, K A 1996 MNRAS 283, L45
\bibitem{ref:9} Barvainis, R E 1990 ApJ 353, 419
\bibitem{ref:10} Begelman, M, Blandford, R, and Rees, M 1984
 Rev Mod Phys 56, 255
\bibitem{ref:11} Blandford, R and Znajek, R 1977 MNRAS 179, 433
\bibitem{ref:12} Carilli, C L, and Barthel, P D 1996 A\&AR 7, 1
\bibitem{ref:13} Capetti, A et al.\ 2000 ApJ 544, 269
\bibitem{ref:14} Carilli, C L, and Barthel, P D 1996 A\&A Reviews 7, 1
\bibitem{ref:15} Chiaberge, M, Capetti, A, and Celotti, A 1999
Astron Astrophys 349, 77
\bibitem{ref:16} Chiaberge, M, Capetti, A, and Celotti, A 2000
 Astron Astrophys 355, 873
\bibitem{ref:17} Cohen, M H, Ogle, P M, Tran, H D, Goodrich, R W, and
\bibitem{ref:18} Djorgovski, S, Weir, N, Matthews, K, and Graham, J R 1991 ApJL
372, 67
Miller, J S 1999 Astron J 118, 1963
\bibitem{ref:19} Ekers, R D, and Simkin, S M 1983 ApJ 265, 85
\bibitem{ref:20} Falcke, H, Malkan, M A, Biermann, P L 1995 A\&A 298, 375
\bibitem{ref:21} Ford, H, and Tsvetanov, Z 1999, in ``The Radio Galaxy M87:
  proceedings of a workshop held at Ringberg Castle" (NY:Springer), 
     Roser and Meisenheimer, eds., p.~278
\bibitem{ref:22} Goodrich, R W, and Miller, J S 1989 ApJ 346, L21
\bibitem{ref:23} Gopal-Krishna, Kulkarni, V K, and Wiita, P J 1996 ApJ 463, L1
\bibitem{ref:24} Hines, D C, and Wills, B J 1993 ApJ 415, 82
\bibitem{ref:25} Impey, C D and Neugebauer, G 1988, Astronom. J, 95, 307
\bibitem{ref:26} Kishimoto, M,  Antonucci, R, Cimatti, A, Hurt, T, Dey, A,
 and ven Breugel, W 2001, ApJ, in press; also astro-ph 0010001
\bibitem{ref:27} Laing, R 1994 Physics of Active Galaxies, ASP Conf Series
 \#54, eds.\ Bicknell et al.
\bibitem{ref:28} Lara, L, Marquez, I, Cotton, W D, Feretti, L,
 Giovannini, G, Marcade, J M, and Venturi, ,T 1999 New Astr Reviews 43, 643
\bibitem{ref:29} Marconi, A, Schreier, E J, Koekemoer, A, Capetti, A,
 Axon, D, Macchetto, D, and Caon, N 2000 ApJ 528, 276
\bibitem{ref:30} Moshir, M. et al.\ Infrared Astronomical Satellite
 Catalogs, 1990, The Faint Source Catalog, version 2.0
\bibitem{ref:31} Ogle, P M, Cohen, M H, Miller, J S, Tran, H D,
 Fosbury, R A E, and Goodrich, R W 1997 ApJ 482, L37
\bibitem{ref:32} Owen, F W, Eilek, J A, and Kassim, N E 2000 AJ 543, 611
\bibitem{ref:33} Punsly, B 2001, monograph in prep
\bibitem{ref:34} Reynolds, C S et al.\ 1996 MNRAS 283, L111
\bibitem{ref:35} Robson, I et al.\ 1993 MNRAS 301, 935
\bibitem{ref:36} Rydbeck, G, Wiklind, T, Wild, W, eckart, A, Genzel, R,
 and Rothermel, H 1993  A\&A 270, L13
\bibitem{ref:37} Sambruna, R M, Chartas, G, Eracleus, M, Mushotzky, R F,
 and Nousek, J A 2000 ApJ 532, L91
\bibitem{ref:38} Sanders, D B, Phinney, E S, Neugebauer, G, Soifer, B T,
 and Matthews, K 1989 ApJ 347, 29
\bibitem{ref:39} Sarazin, C L, Koekemoer, A M, Baum, S A, O'dea, C P, Owen,
 F N, and Wise, M W 1999 ApJ 510, 90
\bibitem{ref:40} Singal, A 1993 MNRAS 262, L27
\bibitem{ref:41} Shaw,, M and Tadhunter, C  1994 MNRAS 267, 589
\bibitem{ref:42} Sikora, M 2001 ASP Conference Series:  Blazar Demographics
and Physics (in press);  also astro-ph 01011381
\bibitem{ref:43} Stockton, A Ridgway, S E, and Lilly, S J 1994 AJ 108, 414
\bibitem{ref:44} Thornton, R J, Stockton, A, and Ridgway, S 1999 AJ 118, 146
\bibitem{ref:45} Tadhunter, C N, Metz, S, and Robinson, A 1994 MNRAS 268, 989
\bibitem{ref:46} Tadhunter, C, et al.\ 2000 MNRAS 313, L52
\bibitem{ref:47} Tsvetanov, Z et al.\ 1998 ApJ 493, L83
\bibitem{ref:48} Vestergaard, M, and Barthel, P D Ap\&SS 205, 135
\bibitem{ref:49} Young, S, Hough, J H, Efstathiou, A, Wills, B J, Axon, D J,
 Bailey, J A, and Ward, M J 1996 MNRAS 281, 1206
\bibitem{ref:50} Wilkes, B J, Schmidt, G D, Smith, P S, Mathur, S, and
 McLoed, K K 1995 ApJ 455, L13
\end{thebibliography}
\end{document}